\documentstyle[prc,aps,epsfig,multicol]{revtex}
\begin{document}
\draft
\title{Antiflow of kaons in relativistic heavy ion collisions}
\author{Subrata Pal, C.M. Ko, Ziwei Lin, and Bin Zhang}
\address{Cyclotron Institute and Physics Department, Texas A$\&$M University,
College Station, Texas 77843-3366}

\maketitle

\begin{abstract}
We compare relativistic transport model calculations to recent data
on the sideward flow of neutral strange $K^0_s$ mesons for Au+Au 
collisions at 6 AGeV. A soft nuclear equation of state is found to
describe very well the positive proton flow data measured in the 
same experiment. In the absence of kaon potential, the $K^0$ flow 
pattern is similar to that of protons. The kaon flow becomes negative 
if a repulsive kaon potential determined from the impulse approximation 
is introduced. However, this potential underestimates the data which 
exhibits larger antiflow. An excellent agreement with the data is obtained
when a relativistic scalar-vector kaon potential, that has stronger 
density dependence, is used. We further find that the transverse 
momentum dependence of directed and elliptic flow is quite sensitive 
to the kaon potential in dense matter.
\end{abstract}


\begin{multicols}{2}
 
The study of hadron properties in hot and dense medium created in 
relativistic heavy ion collision is of considerable interest. Of 
particular importance is the in-medium modification of kaon properties
as it is related to chiral symmetry restoration \cite{Brow1}
and neutron star properties \cite{Brow2}. Since the work of Kaplan 
and Nelson \cite{Kapl} on the possibility of kaon condensation in 
dense matter, considerable effort has been devoted to the understanding
of kaon properties in dense matter from heavy ion collisions
[4-10]. All these investigations
have lead to a general consensus that kaons feel a weak repulsive
potential while antikaons feel a strong attractive potential \cite{Ko}; the
latter is responsible for $K^-$ condensation in neutron stars. 
Since kaons are primarily produced at an early hot and dense stage
in heavy ion collision, a promising tool to search for kaon in-medium
properties is via pressure-induced collective flow effects. The 
strength of the flow depends on the nuclear equation of state (EOS),
the pressure gradient developed, and the kaon potential or dispersion
relation in dense matter. Indeed, Li {\it et al.} \cite{Li5}, based on a 
relativistic transport model, have found that the repulsive kaon potential
in nuclear medium tends to deflect kaons away from nucleons
leading to an anticorrelated flow. On the other hand, antikaons
were found \cite{Li11} to have a similar flow as that of nucleons due to
their attractive potential.

The theoretical predictions of exploiting the transverse flow to study
the kaon self-energy in the medium have attracted much attention 
experimentally. The FOPI Collaboration at SIS/GSI has found 
\cite{Rit,Best,Croc}
for the reaction Ni+Ni at $E_{\rm beam}=1.93$ AGeV a very small antiflow
for kaon, i.e., the kaon average transverse momentum in the reaction plane
as a function of rapidity is almost zero around midrapidity. The kaon 
sideward flow for this reaction, and especially for a heavier system
Ru(1.69 AGeV)+Ru, was found to be anticorrelated (correlated) with that
of protons at low (high) transverse momenta. The azimuthal angular
distributions of kaons measured by the KaoS Collaboration \cite{Shin}
in Au+Au collisions at 1 AGeV indicate that kaons are emitted preferentially
perpendicular to the reaction plane. Also large $K^-$ production
cross-section was observed by this group in Ni+Ni collisions at 
$E_{\rm beam} = (0.8-1.8)$ AGeV \cite{Bart}. All these experimental
findings at beam energies of $1-2$ AGeV clearly suggest the existence 
of an in-medium repulsive kaon-nucleon potential and a strong attractive
antikaon-nucleon potential; the latter is also supported by $K^-$
atomic data \cite{Frie}.

Flow analysis for $K^+$, $K^-$, and $K^0_s$ have been performed
by several collaborations at AGS/BNL in Au+Au collision at beam energies of 
$2-12$ AGeV \cite{Ogil,Vol98}. Most of these experiments are confined to 
near-central collisions where nearly vanishing kaon transverse flow
versus rapidity was observed. Several suggestions put forward about the 
origin of vanishing flow are, for example, isotropic production
of kaons in hadron-hadron collisions \cite{Ogil}; the colliding hadrons
have opposite rapidities \cite{Dav}; cancellation between the
negative flow at low transverse momenta and the positive flow at high
transverse momenta \cite{Bali10}.

Recently, the E895 Collaboration at AGS/BNL has measured \cite{Chu}
the directed flow of neutral strange $K^0_s$ mesons in 6 AGeV Au+Au
collisions for central and mid-central events. The $K^0_s$ was found to
have a considerable antiflow relative to that of the proton observed in 
the same experiment \cite{Liu}. Using a relativistic transport model (ART)
\cite{art} for heavy ion collisions we investigate in this paper
if the appreciable kaon antiflow can be explained by the repulsive kaon 
potential in dense nuclear matter. We shall demonstrate that the typical
relativistic scalar-vector mean field potential can indeed reproduce
the data. We also predict that the flow anisotropies is sensitive
to the kaon potential and is thus a useful probe of the EOS in high 
density matter.

In the ART model employed here, the nucleon mean field is parameterized by 
the usual Skyrme-type with a soft EOS corresponding to an incompressibility of 
$K=210$ MeV and a stiff EOS with $K=380$ MeV at normal nuclear matter density 
of $\rho_0=0.16$ fm$^{-3}$. In this model, the imaginary part of the kaon 
self-energy is approximately treated by kaon-hadron scatterings and 
its real part is given by the mean-field potential. 
The model has been used to study many aspects
of heavy ion collisions at the AGS energies \cite{Bali10,art1,Bali30}.
In the original ART model, kaon production in hadron-hadron scatterings
is instantaneous. In the present study, we introduce a formation time
of $\tau^K_{\rm fo}=1.2$ fm/c. 

Various approaches have been adopted
to evaluate the kaon potential in dense matter \cite{Li5,Li9}. 
In the present study, we shall use the two commonly used forms that have 
been quite successful in explaining several experimental data. One of these
is based on the kaon dispersion relation determined from the 
kaon-nucleon scattering length using the impulse approximation:
\begin{equation}
w_K\left({\bf p}, \rho_b \right) = \left[ m_K^2 + {\bf p}^2 - 4\pi 
\left(1 + \frac{m_K}{m_N} \right) {\bar a}_{KN} \rho_b \right]^{1/2} ,
\end{equation}
where $m_K$ and $m_N$ are the kaon and nucleon bare masses, $\rho_b$ is
the baryon density and ${\bar a}_{KN} \approx -0.255$ fm is the
isospin-averaged kaon-nucleon scattering length. The kaon potential
in nuclear medium can then be defined as 
\begin{equation}
U_K\left({\bf p},\rho_b \right) = w_K\left({\bf p},\rho_b \right) 
- \left( m_K^2 + {\bf p}^2 \right)^{1/2} .
\end{equation}
This yields a repulsive potential of 30 MeV at $\rho_0$ for kaon at 
zero momentum. The other form of kaon potential used here is the 
scalar-vector potential determined from the chiral Lagrangian and can be
written as
\begin{equation}
w_K\left({\bf p}, \rho_b \right) = \left[ m_K^2 + {\bf p}^2 - a_K\rho_S
+ (b_K\rho_b)^2 \right]^{1/2} + b_K\rho_b ~ ,
\end{equation}
where $b_K = 3/(8f^2_\pi) \approx 0.333$ GeV fm$^3$, $\rho_S$ is the 
scalar density. If only the Kaplan-Nelson term, which is related
to the $KN$ sigma term due to explicit chiral symmetry breaking,
is considered, then the
parameter $a_K$ which determines the strength of the attractive kaon potential
is $a_K = \Sigma_{KN}/f_\pi^2$. Since the exact value of $\Sigma_{KN}$ and
the magnitude of the higher order corrections are poorly known, we take
a different approach where $a_K \approx 0.173$ GeV$^2$ fm$^3$ is determined 
by fitting the kaon repulsive potential of $U_K(\rho_0)= + 30$ MeV as obtained 
in the impulse approximation.

Let us now make a systematic study of the directed flow of particles measured
by the E895 Collaboration for the Au+Au collisions at $E_{\rm beam} = 6$ AGeV
\cite{Chu,Liu}. Before we confront with the $K^0_s$ flow data, it is 
instructive to find how far the ART model can describe the observed proton 
flow \cite{Liu}. 
Calculations are performed for impact parameters $b=5-7$ fm, since
the events analyzed for the proton flow originate from this domain \cite{Liu}.
Figure 1 shows for protons the mean transverse momentum in the reaction plane 
$\langle p_x \rangle$ as a function of rapidity $y$ (scaled to the beam
rapidity). The data reveals a large proton flow with a slope of 
$F = d\langle p_x \rangle /d (y/y_{c.m.}) \sim 180$ MeV/c at midrapidity.
The ART model with a soft nuclear equation of state (solid line) provides 
a good agreement with the data over a large rapidity range. On the other 
hand, 
{\centerline{
\epsfxsize=7.5cm
\epsfysize=6.5cm
\epsffile{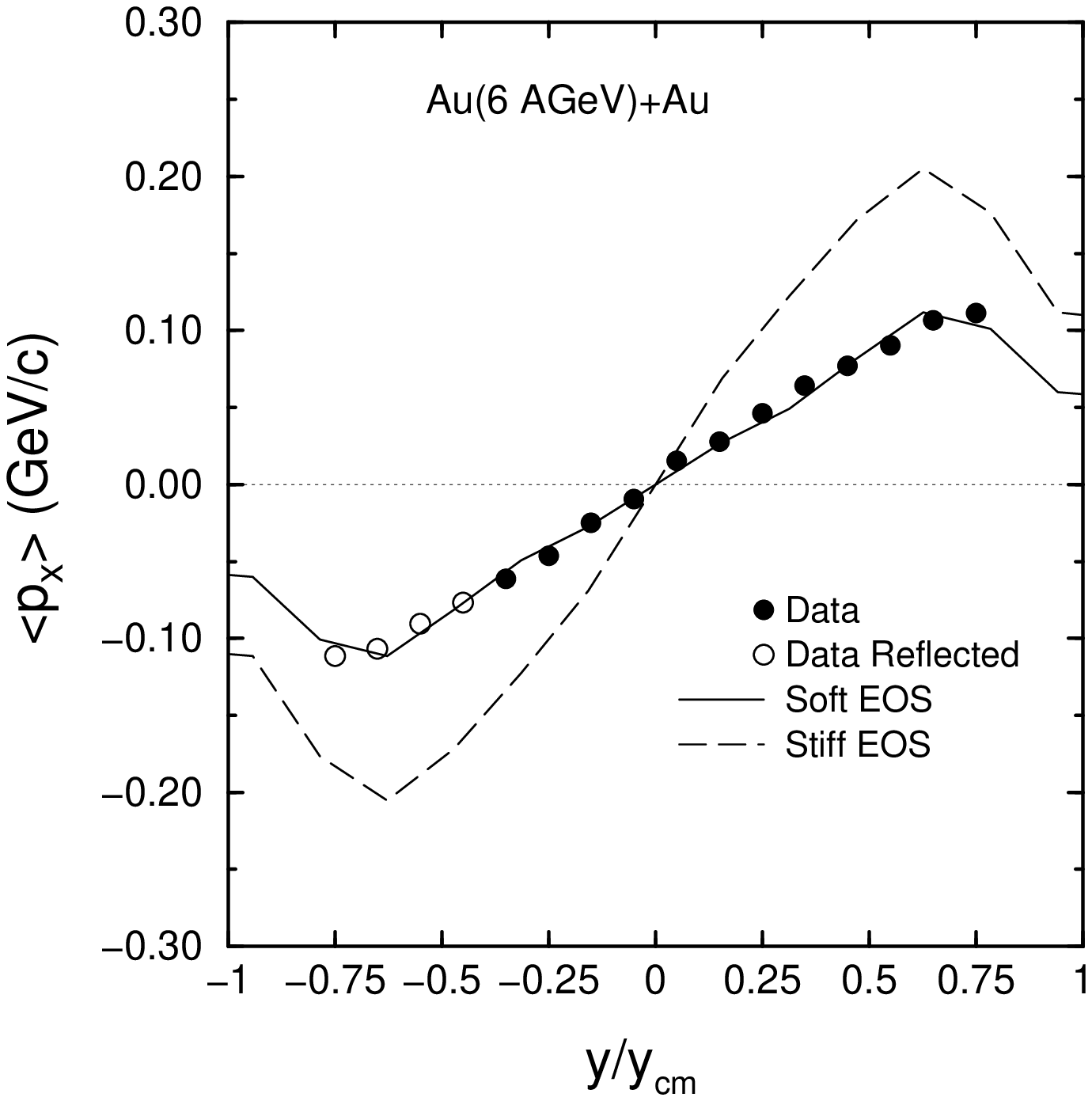}
}}
\vspace{-.5cm}

{\small{
FIG. 1. The proton directed transverse flow $\langle p_x\rangle$ as a function
of normalized rapidity $y/y_{cm}$ in Au+Au collisions at $E_{\rm beam}=6$ AGeV. 
The ART model calculations for a soft (solid line) and stiff (dashed line) 
nuclear EOS at an impact parameter $b=5-7$ fm are compared with the E895 
data (filled circles).}}
\vspace{.7cm}

\noindent
the stiff EOS (dashed line) results in a much stronger proton flow compared 
to the data. The enhanced pressure gradient for the stiff EOS 
provides a larger driving
force for the expansion of the system and thereby a stronger collective flow.
Henceforth, we restrict only to the soft nuclear EOS when comparing with the
$K^0_s$ antiflow data measured in the same experiment.

In Fig. 2, the ART model predictions of the sideward kaon flow as a function of
rapidity are compared with the E895 data of $K^0_s$ mesons reconstructed from
charged pion decays \cite{Chu}. Since the antikaon $\bar K^0$ has a 
contribution
of $\stackrel{<}{\sim} 10\%$ in the data sample at AGS energies, the observed
$K^0_s$'s are therefore primarily from neutral kaons $K^0$'s. It is thus 
justified
to study the collective flow of $K^0$ mesons in our model. In contrast to 
the protons, 
a pronounced in-plane antiflow for the $K^0_s$ is observed in the data with a
slope of $F\simeq -127 \pm 20$ MeV/c. The $K^0_s$ flow data corresponds
to central and mid-central events with $b\stackrel{<}{\sim} 7$ fm and for
transverse momentum $p_t \leq 700$ MeV/c \cite{Chu}. We therefore restrict 
our calculations to these cut-off values. As is evident from Fig. 2, in the
absence of kaon mean field, the $K^0$'s have a flow pattern similar to 
that of the nucleons. This is not surprising as most of 
the $K^0$'s are produced in the early compression stage of the reaction at 
time $t\stackrel{<}{\sim} 5$ fm/c and the rescattering of the kaons with 
the nucleons in the dense matter thus causes them to flow in the 
direction of the nucleons.

We shall now demonstrate the effect of inclusion of kaon potential
on the $K^0$ flow. Using the kaon potential determined from the
impulse approximation, the $K^0$'s are repelled from the nucleons resulting
in antiflow with respect to the nucleons (see Fig. 2). However, the
$\langle p_x \rangle$ for kaon is found to underestimate the data
as in Ref. \cite{bzhang}.
{\centerline{
\epsfxsize=7.7cm
\epsfysize=6.5cm
\epsffile{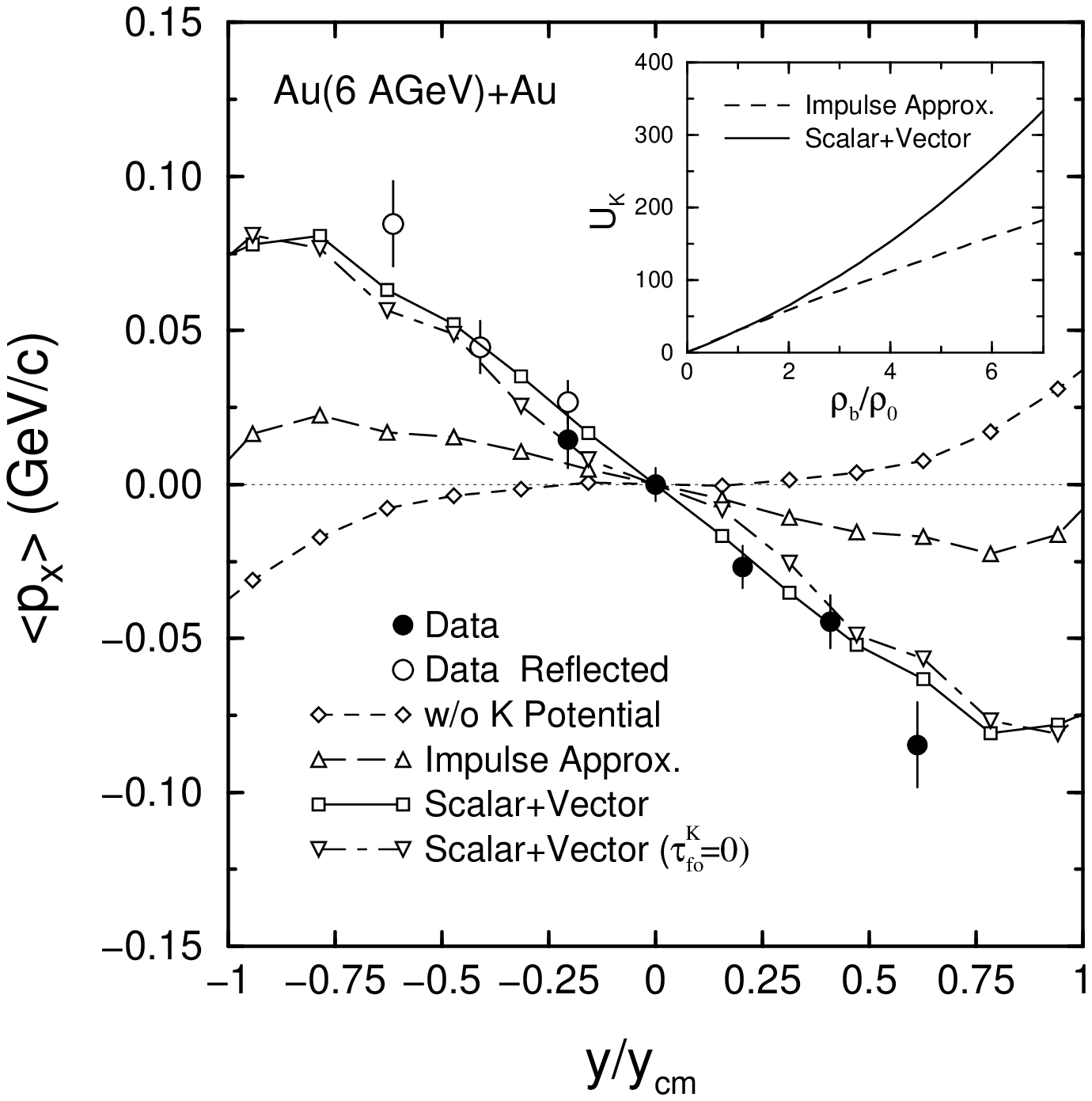}
}}
\vspace{-.5cm}

{\small{
 FIG. 2. The kaon directed transverse flow $\langle p_x\rangle$ as a function
of normalized rapidity $y/y_{cm}$ in Au+Au collisions at $E_{\rm beam}=6$ AGeV. 
The filled circles are the E895 data while the other curves correspond to ART 
model calculations ($b\stackrel{<}{\sim} 7$ fm and $p_t \leq 700$ MeV/c)
without kaon mean field potential and with kaon dispersion relations obtained 
from impulse approximation and scalar-vector potential. For the latter kaon 
potential, results with zero kaon formation time $\tau_{\rm fo}^K$ is also shown.
The inset shows the density dependence of the kaon potential $U_K$ at zero 
momentum obtained in the impulse approximation and scalar-vector potential.}} 
\vspace{.7cm}

\noindent
In fact, theoretical calculations 
\cite{Brow23} have shown that the simple impulse
approximation also underestimates experimental kaon-nucleus scattering data. 
The scalar-vector potential has clearly a stronger density 
dependence compared to the impulse approximation (see inset of Fig. 2)
since the vector potential dominates over the scalar potential at
densities above the normal nuclear matter value. The kaon flow predicted
from the scalar-vector potential is found to have a strikingly good 
agreement with the $K^0_s$ flow data as seen in Fig. 2. 
Although the final kaon flow is opposite to the proton flow, the primordial
kaons flow with the nucleons up to the maximum compression stage of the
collision, and their $\langle p_x\rangle$ values are nearly identical 
irrespective of the kaon potential employed. This stems from the fact that 
at the early stage of the collision the density gradient inside the 
participant matter is rather small and as a consequence the force acting 
on the kaon during its propagation is negligible. The potential 
(or baryonic density) gradient is substantial only during the expansion 
of the system so that the kaon potential could
then effectively repel the $K^0$'s away from the nucleons. This suggests
that at the AGS energies, kaon flow may provide information about the
kaon dispersion relation for densities below $\sim 4\rho_0$. Since the kaons 
produced in the ultradense stage mostly travel with nucleons, the kaon flow
is expected to be sensitive to its formation time. This effect is illustrated
in Fig. 2 for the scalar-vector potential where the kaon formation time 
$\tau_{\rm fo}^K$ is set to zero. The early production of $K^0$ then 
leads to larger rescattering in the initial stage of the reaction. 
Consequently, the antiflow of $K^0$'s is suppressed especially near the 
midrapidity. It is worth mentioning that the scalar-vector potential has 
been successful in reproducing several data at GSI energies of $1-2$ AGeV 
\cite{Li9}. 

Additional insight into the origin of the transverse flow may be gained by
considering the azimuthal angular distribution $dN/d\phi$ of the kaons at
midrapidity by means of the Fourier expansion \cite{Vol97,Posk}
\begin{equation}
\frac{dN}{d\phi} = v_0 \left[ 1 + 2v_1 \cos\phi + 2v_2 \cos(2\phi) + \cdots \right] .
\end{equation}
Here $v_0$ is the normalization constant, and the Fourier coefficients 
$v_1 = \langle \cos\phi \rangle = \langle p_x/p_t\rangle$ represents the
directed flow and
$v_2 = \langle \cos(2\phi) \rangle = \langle (p_x^2-p_y^2)/p_t^2\rangle$,
referred to as the elliptic flow \cite{Olli}, reflects the azimuthal asymmetry 
of the emitted particles. While $v_2>0$ indicates in-plane enhancement, $v_2<0$
suggests a squeeze-out orthogonal to the reaction plane.

In Fig. 3 we depict the $p_t$-dependence of $v_1$, dubbed as differential flow,
for midrapidity $K^0$ ($|y/y_{c.m.}|< 0.3$) without any kaon mean
field and with the inclusion of scalar-vector kaon potential. The results
are for semicentral ($b=6$ fm) Au+Au collisions at 6 AGeV which can
describe equally well the flow data. In absence of kaon potential, $v_1$ is
found to be almost zero at low $p_t$ but has a large
positive value at high $p_t$. The low-$p_t$ kaons are produced mostly in the
relatively dilute region of the central expanding hadronic matter and are
preferentially emitted in the antiflow direction away from the spectators.
Consequently, they freeze out quite early in the reaction leading to a small
in-plane flow. In contrast, kaons with large final $p_t$ originate mainly inside
the baryon-rich participant zone. These kaons are likely to suffer multiple 
rescatterings with the participant baryons and with the spectators. Hence 
the transverse momenta of these kaons are pushed to higher values with the
final flow direction (and magnitude) similar to those of the nucleons. The 
differential flow for midrapidity nucleons seen here is similar to that 
observed experimentally \cite{Barr}.

In presence of repulsive scalar-vector potential for $K^0$, the 
differential flow
of kaons with $p_t \leq 1.2$ GeV/c becomes negative while kaons with higher
$p_t$ remains positive. The large negative $v_1$ for low-$p_t$ kaons follows
from the fact that these kaons are formed near the surface of the participant 
matter and thus feel a 
pronounced repulsive (kaon) potential gradient away from the baryons. Also
the force acting on a kaon being inversely proportional to energy, the low
energy  kaons are strongly repelled from the baryons. Since the high-$p_t$
kaons are produced deep inside the baryon-rich medium they are less repelled
by the kaon potential both due to smaller density gradient and eventually 
because of their larger energy. These kaons thus remain to flow in the same 
direction as the baryons. Note that the $p_t$-cut in the E895 data for 
$K^0_s$ flow ($p_t \leq 0.7$ GeV/c) leads to enhanced 
{\centerline{
\epsfxsize=7.5cm
\epsfysize=6.5cm
\epsffile{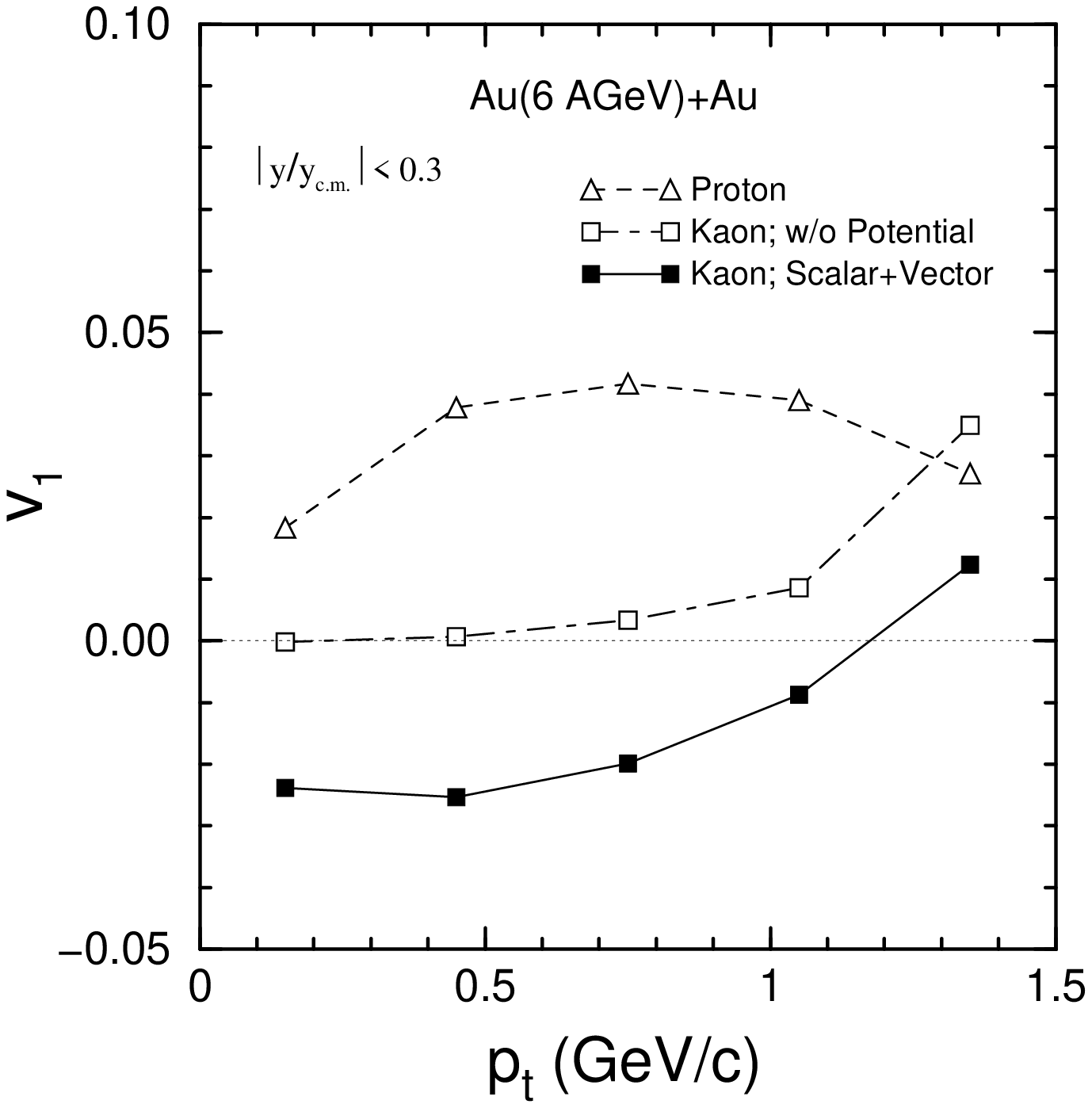}
}}
\vspace{-.5cm}

{\small{
FIG. 3. The transverse momentum dependence of the directed flow $v_1$ for 
midrapidity protons and kaons without and with a kaon scalar-vector 
potential. The results are for Au(6 AGeV)+Au collisions at an impact 
parameter of 6 fm.}}
\vspace{.7cm}

\noindent
antiflow for the $K^0_s$ than that when integrated over the 
entire $p_t$ spectra.

In Au(6 AGeV)+Au collisions the passage time of projectile and target spectator
is given by $t\approx 2R/(\gamma v) \approx 6.4$ fm/c. During
this time, the spectators prevent the participating hadrons from developing
an in-plane flow. Therefore the hadronic matter is initially squeezed-out
preferentially orthogonal to the reaction plane. The sign and magnitude of 
elliptic flow $v_2$, apart from the passage time, is also determined by the
transverse expansion time of the participant matter. In the later stages
of the reaction, however, the geometry of the participant region is such that
in-plane emission is favored resulting in a positive $v_2$ \cite{Olli,Sorg}. 
Most of the kaons at the AGS energies are produced during the compression 
stage at
times less than $\sim 5$ fm/c and thus would be shadowed by the spectators.
Therefore kaons may serve as a sensitive probe to the EOS in the ultradense
matter. Indeed, in peripheral and semicentral Au+Au collisions at $\sim 1$ 
AGeV, preferential emission of $K^+$ meson perpendicular to the reaction 
plane has been
observed \cite{Shin}. This may be interpreted as an evidence of repulsive kaon 
potential in the medium \cite{Wang}.

In Fig. 4, the elliptic flow $v_2$ as a function of $p_t$ is shown for protons
and kaons using the same cuts as in Fig. 3. The proton $v_2$ gradually 
increases with $p_t$ and have positive values even for small $p_t$.
The low-$p_t$ particles generally freezes out in the compression stage 
orthogonal to the reaction plane due to spectator shadowing,
while the high-$p_t$ particles undergo multiple rescattering. The experimental
observation \cite{Pike} for proton $v_2$ indicates a transition from 
squeeze-out
to a preferential in-plane emission with increasing beam energies; the
transition occurs at $E_{\rm beam} \sim 6$ AGeV. The ART model \cite{Bali30}
and the UrQMD model \cite{Soff} can describe the data consistently with a 
stiff nuclear EOS ($K=380$ MeV); 
the $v_2$ for protons
{\centerline{
\epsfxsize=7.5cm
\epsfysize=6.5cm
\epsffile{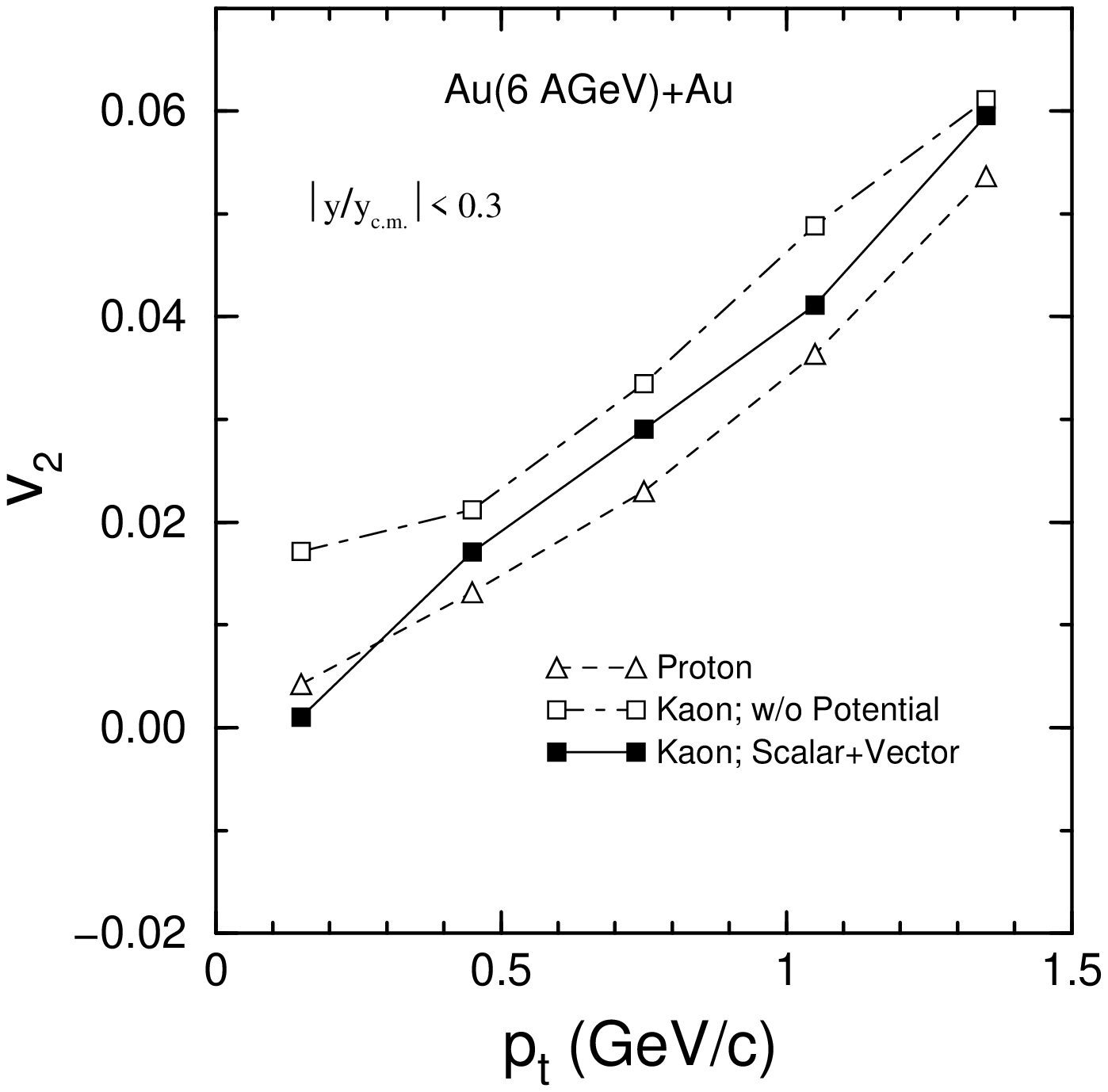}
}}
\vspace{-.5cm}

{\small{ FIG. 4. Same as Fig. 3, but for the elliptic flow $v_2$.}}
\vspace{.8cm}

\noindent
in the ART model is however insensitive to the EOS for 
$E_{\rm beam} \geq 6$ AGeV. This is in contrast to the prediction of Ref. 
\cite{Dani} where a softening of the EOS with incident energy is deduced.

The elliptic flow for $K^0$ in absence of any kaon potential is found to be 
slightly larger than that of the protons. This is caused by smaller shadowing
by the spectators because of the time delay in their production and due to
small $K-N$ total cross-section of about 10 mb. With the inclusion of the
scalar-vector potential, the $v_2$ for kaons is reduced.
An explanation to this is as follows. The geometrical shape of the participant 
region at the initial stages entails a kaon repulsive potential
gradient that is larger in the direction perpendicular to the reaction plane 
than parallel to the plane. Consequently, kaons in its mean field escape more
freely perpendicular to the reaction plane. This effect is more pronounced
for low-$p_t$ kaons produced near the surface of the participant region. We
however find that at the AGS energies considered here, the sensitivity of the
squeeze-out to the kaon dispersion relation is smaller compared to that
observed at GSI energies \cite{Li6,Wang,Shin}.

In summary, recent data on the sideward flow of neutral strange $K^0_s$ meson
in 6 AGeV Au+Au collisions have been compared to ART model calculations.
The directed flow data for $K^0_s$ is opposite to that of protons and reveals
a large antiflow. A soft nuclear EOS is found to reproduce very well the 
in-plane proton flow data at the same beam energy. We have used two different
kaon dispersion relations: the impulse approximation where the mean field has
a linear density dependence, and the typical relativistic scalar-vector 
potential. The $K^0$ mesons which flow with the nucleons in absence of a kaon
mean-field potential exhibit an antiflow for the kaon potential derived from
impulse approximation. However, the antiflow is found to underestimate the 
data.
An overall good agreement with the data is achieved for the scalar-vector 
potential that has been equally successful in explaining the kaon data at
SIS/GSI energies. Since the vector potential dominates at high densities, the
scalar-vector potential becomes more repulsive than that in the impulse
approximation. The $K^0$ flow at AGS energies is sensitive to the kaon potential
at densities smaller than about four times the normal nuclear matter value 
as the kaon flow is essentially determined during the expansion stage of the
collision. At higher densities reached at the early stages, the kaons generally
flow with the nucleons and their flow magnitude is thus determined by the nuclear EOS.
Kaons with small transverse momentum have a large negative differential flow
due to stronger repulsion by the kaon potential, while positive kaon differential 
flow is a consequence of strong positive baryonic flow. In the semicentral
collisions about midrapidity, the kaons are found to have an in-plane 
elliptic flow. The magnitude of this flow is slightly larger than that of
the protons since the kaons suffer smaller shadowing. In presence of 
scalar-vector kaon potential, the kaon elliptic flow is decreased. This 
suggests that analysis of directed flow in conjunction with elliptic flow 
of kaons may provide useful information on the kaon dispersion relation
in dense matter.

\bigskip

This work was supported in part by the National Science Foundation under 
Grant No. PHY-9870038, the Welch Foundation under Grant No. A-1358,
and the Texas Advanced Research Program under Grant No. FY99-010366-0081.

\end{multicols}
\end{document}